\documentclass[]{interact}

\usepackage{epstopdf}
\usepackage{graphicx}
\usepackage{subcaption}
\usepackage{hyperref}
\usepackage{cleveref}
\usepackage{tabularx}
\usepackage{makecell}

\usepackage[numbers,sort&compress]{natbib}
\bibpunct[, ]{[}{]}{,}{n}{,}{,}
\renewcommand\bibfont{\fontsize{10}{12}\selectfont}

\usepackage{interval}
\usepackage{xparse,mathtools}
\ExplSyntaxOn

\NewDocumentCommand \vecrow { s o m }
 {
  \IfBooleanTF {#1}
   { \vectaux*{#3} }
   { \IfValueTF {#2} { \vectaux[#2]{#3} } { \vectaux{#3} } }
  ^\top
 }

\DeclarePairedDelimiterX \vectaux [1] {\lbrack} {\rbrack}
 { \, \dbacc_vecrow:n { #1 } \, }

\cs_new_protected:Npn \dbacc_vecrow:n #1
 {
  \seq_set_split:Nnn \l_tmpa_seq { , } { #1 }
  \seq_use:Nn \l_tmpa_seq { \enspace }
 }
\ExplSyntaxOff

\theoremstyle{plain}

\theoremstyle{definition}

\theoremstyle{remark}

\newcommand{\obs}{b}
\newcommand{\Obs}{\mathcal{B}}
\newcommand{\bb}{\mathbb}
\newcommand{\dif}{\, \mathrm{d}}
\newcommand{\cDisk}{D}
\newcommand{\oo}{\infty}
\newcommand{\abs}[1]{\lvert #1 \rvert}
\newcommand{\norm}[1]{\left\| #1 \right\|}

\newcommand{\opMSE}{\operatorname{MSE}}
\newcommand{\opTV}{\operatorname{TV}}
\newcommand{\opT}{\operatorname{T}}
\newcommand{\rhoMSE}{\rho_\text{MSE}}
\newcommand{\rhoTV}{\rho_\text{TV}}
\newcommand{\rhoT}{\rho_\text{T}}

\begin{document}

\title{Probability-Based Collision Risk Evaluation of Trajectories for Optimal Control Problems with Moving Obstacles}

\author{
\name{Florian Steppich\textsuperscript{a}\thanks{CONTACT Email: florian.steppich@unibw.de} and Matthias Gerdts\textsuperscript{a}}
\affil{\textsuperscript{a}Institute of Applied Mathematics and Scientific Computing, Department of Aerospace Engineering, University of the Bundeswehr Munich, 85577 Neubiberg, Germany}
}

\maketitle

\begin{abstract}
This paper presents a method for approximating occupancy distributions, a common probabilistic representation used in trajectory planning for autonomous vehicles operating in uncertain environments. The proposed method employs B-spline surfaces in conjunction with regularization techniques to ensure smoothness and differentiability. Interpreting the resulting approximation as an intensity function of a Poisson random field enables seamless integration of multiple occupancy distributions into a coherent risk representation, while retaining compatibility with the semantics of probability distributions.

The method is tailored to optimal control problems (OCPs), where solvers benefit from gradient and higher-order derivative information. It also preserves information about the spatial and temporal structure of occupancy distributions, which is critical for handling dynamic obstacles. We demonstrate the feasibility of the approach in a trajectory planning scenario involving autonomous vehicles and moving obstacles with time-varying uncertainty.

By enabling controlled risk acceptance, our formulation extends beyond conservative "no-collision" strategies and allows for admissible trajectories that would otherwise be ruled out. In our application scenario, we report a reduction in solver computational load of up to 50\% through proper tuning of the regularization.
\end{abstract}


\section{Introduction}
This paper addresses the problem of planning collision-free trajectories for autonomous ground vehicles operating in dynamic environments populated by moving obstacles such as cars, trucks, bicycles, and pedestrians. We propose a method for generating a time-varying map that can serve as a risk-aware guide for trajectory planning.

Trajectory planning methods in dynamic environments typically address two interrelated problems: first, modeling the environment around the ego vehicle, and second, identifying a viable trajectory to navigate the vehicle safely in that environment.

The authors \cite{wang2024} represent the environment by constructing a map describing the severity of a collision in a certain area. Each obstacle is approximated by a simple geometric shape (e.g., circles for pedestrians, rectangles for cars), generating a cost field with constant height that smoothly decays to zero at the edges. Multiple obstacles are handled through superposition of these fields. The height of each field is scaled to reflect the severity of a collision -- for example, collisions with humans are penalized more than with vehicles. In an earlier study, \cite{wang2022} adopted multivariate B-splines to model the map, which allows for smoother and more flexible representations. Their trajectory planning framework consists of a two-step optimal control process: the first stage proposes a feasible candidate trajectory, while the second selects the one with minimal control effort.

In the work \cite{boonyarak2014} potential fields are constructed from attractive forces toward the goal and repulsive forces from both static and moving obstacles. The magnitude of each force is based on the relative position and velocity between the ego vehicle and the surrounding objects. The vehicle's trajectory is planned incrementally, minimizing the net force acting on it at each time step.

The authors of \cite{ji2017} define a potential field that encourages progress along a reference trajectory, such as a road lane. Obstacles are represented by a repelling potential field in order to maintain a safe distance and make lane switching a considerable option. The method solves a series of multi-constrained model predictive control (MMPC) problems to compute steering angles. Notably, the vehicle's speed is not controlled in this framework; longitudinal motion (acceleration, braking) is managed by a separate emergency braking component. Obstacle motion is accounted for by updating their position at each MMPC iteration.

In \cite{ge2002} a time-dependent potential field gets constructed such that it incorporates both attraction to the goal and repulsion from moving obstacles. The formulation allows for either soft approaches -- reducing both relative distance and velocity -- or hard approaches -- reducing only distance. The authors also propose solutions to two failure scenarios: when an obstacle persistently blocks access to the goal, and when the goal lies within a region dominated by the obstacle's repulsion.

Occupancy grids, where space is discretized into cells, each assigned a probability of being occupied, is explored in \cite{elfes1989}. These probabilities are derived from probabilistic sensor models and fused using a maximum a posteriori estimator. A path is then planned using A\* search, which takes both distance and occupancy probability into account.

Obstacles get avoided in \cite{weng2024} by maximizing the distance between the ego vehicle and potential collisions while minimizing control effort. This is accomplished via a two-step approach: a genetic algorithm first generates a coarse, collision-free trajectory, which is then refined using sequential quadratic programming to satisfy trajectory constraints.

The authors of \cite{dutoit2011} model the positions of the ego vehicle and obstacles using either dependent, or independent Gaussian distributions, to account for or ignore vehicle interactions. This formulation allows the probability of collision to be computed analytically. The authors suggest using this collision probability as part of a chance-constrained path-planning formulation.

In \cite{hakobyan2019} the authors employ the conditional value-at-risk (CVaR) to measure the risk of a proposed trajectory and consequently refine the trajectory to minimize it. In this setup, obstacles are modeled by convex polytopes with a linear transformation approximating the movement between successive time steps. The work \cite{majumdar2020} discusses the CVaR and potential pitfalls in its application. The authors further relate it to other risk measures such as the Value at Risk (VaR), expected cost, worst case, and entropic risk.

In summary, existing approaches address the uncertainty introduced by moving obstacles using simplified or static probability distributions. However, the actual motion of these obstacles is typically not updated or incorporated during each planning iteration, which limits the responsiveness of the methods.

Path-planning methods based on artificial potential fields tend to rely on local force evaluations to evolve the trajectory. This could lead to suboptimal solutions, as global information is neglected. Furthermore, the functions used to construct these fields frequently involve non-differentiable components (e.g., sign or absolute value functions), which can hinder the performance of optimization algorithms that rely on information of (higher) derivatives.

Additional constraints such as smoothness or limiting curvature to improve passenger comfort, can be cumbersome to express within the potential field framework. Similarly, methods that rely on discretized state spaces (e.g., occupancy grids) require extra processing to generate smooth and physically feasible trajectories.

Finally, human drivers often tolerate a small amount of risk in real-world driving scenarios. Yet, most reviewed approaches adopt a strict "no-collision" mindset. This can result in overly conservative behavior, where the trajectory planner might fail to find a feasible trajectory and therefore halts the vehicle altogether.

To address the limitations identified in the mentioned approaches, we propose a modeling framework that enables time-dependent quantification of collision risk for trajectory planning. This risk arises from multiple moving obstacles, each described by a time-varying occupancy distribution, i.e., the probability distribution over their occupiable future locations. The resulting risk function is intended to serve both as part of the objective function and constraints in optimal control problems (OCPs). A key requirement of the risk function is that it must be sufficiently smooth to allow the usage of solvers relying on gradient and higher derivative information. To enforce this requirement, we approximate the occupancy distributions using non-negative, continuous functions, while considering regularization requirements.

At the beginning of each path planning iteration, we assume updated estimates of obstacle states -- such as position and heading -- are available. The specifics of the guiding model predictive control (MPC) scheme that handles the path planing over a longer time horizon, and the uncertainty propagation method used to generate the occupancy distributions (e.g., \cite{xiu2010,meng2021,steppich2023}) are beyond the scope of this work.

Although the proposed risk model is developed with OCP applications in mind, it is also applicable to discrete planning frameworks, such as graph-based (e.g., Dijkstra, A\*) or sampling-based algorithms (e.g., PRM, RRT), which may benefit from a continuous and probabilistic interpretable collision measure of risk.

Please note, that we use the term risk in a broad sense, until we arrive at a precise definition at the end of \cref{secApplicationPrerequisitesInterpretation}.

This paper's organization is as follows: \cref{secApplicationPrerequisites} will formally describe the environment, the ego vehicle, and the considered obstacles in this work. The occupancy distributions and three interpretation perspectives are discussed in \cref{secApplicationPrerequisitesInterpretation}. This section concludes with our definition of risk as an intensity map. This term and its relation to the Sobolev space, and B-Spline surfaces in particular, is further addressed in \cref{secApplicationPrerequisitesBSplineSurface}. Subject of \cref{secApplicationApproximationProblem} is the approximation and regularization of the occupancy distribution by an intensity map. \Cref{secApplicationEvaluation} introduces the metrics used to evaluate the approximation (\cref{secApplicationEvaluationApproximationOfOccupancyDistribution}) and the effect on the generation of trajectories based on these intensity maps in \cref{secApplicationOCP}. Finally, \cref{secApplicationConclusion} concludes this work followed by a brief outlook (\cref{secApplicationOutlook}).

\section{Prerequisites} \label{secApplicationPrerequisites}
Let $t$ be the time in the compact interval $\interval{t_0}{t_f} \eqcolon T$, i.e., the temporal domain (e.g., of one MPC iteration).
We approximate an obstacle $\obs$ in the set of all obstacles $\Obs$ by a point mass located at the Cartesian coordinates $\vecrow{\obs_x (t), \obs_y (t)} \in \bb{R}^2$ on the plane, with a heading angle $\obs_\psi (t) \in \bb{R}^2$ measured against the $x$-axis. An obstacle is therefore fully captured by the state 
\[
  \obs(t) = \begin{bmatrix}
      \obs_x (t) \\ 
      \obs_y (t) \\ 
      \obs_\psi (t)
  \end{bmatrix}.
\]

\begin{figure}
    \centering
    \includegraphics[page=2,width=0.5\linewidth]{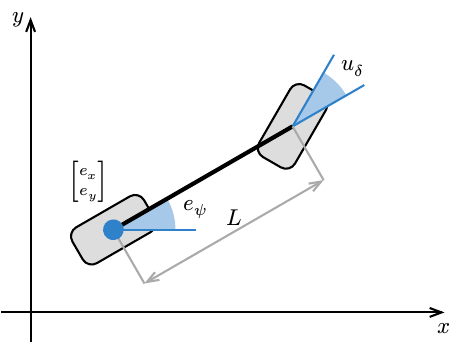}
    \caption{Single-Track Kinematic Model}
    \label{figApplicationModelSingleTrack}
\end{figure}

In a similar manner, the ego vehicle is approximated by the single-track kinematic model. This model is suitable for ground vehicles such as bicycles, motorcycles and cars (cf. \cite{siegwart2011, rill2011}). For the ego vehicle $e$, let $L$ be the distance between the front and rear axle of the vehicle, the Cartesian coordinates $\vecrow{e_x (t), e_y (t)} \in \bb{R}^2$ as the location of the rear axle in the plane, $e_\psi (t) \in \bb{R}$ the heading with respect to the $x$-axis, and $u(t) := \vecrow{u_v (t), u_\delta (t)}$ the control input composed of the linear velocity $u_v (t)$ in the direction of the heading, and $u_\delta (t)$ the steering angle.
\Cref{figApplicationModelSingleTrack} depicts the schematics of the ego vehicles model. 
The dynamics $\dot{e}(t) = g(e(t), u(t))$ are captured by the ordinary differential equations (ODEs)
\begin{equation}
    \begin{aligned}
        \dot{e}_x (t) &= u_v (t)\, \cos(e_\psi (t)),\\
        \dot{e}_y (t) &= u_v (t)\, \sin(e_\psi (t)),\\
        \dot{e}_\psi (t) &= \frac{1}{L} u_v (t)\, \tan(u_\delta (t)).
    \end{aligned}
  \label{eqApplicationODESingleTrackKinematic}
\end{equation}

The simplification of using a point mass neglects critical properties like the physical dimensions of the vehicles. This issue will be addressed as part of the OCP's objective function. 

\begin{figure}
    \centering
    \includegraphics[page=1,width=0.5\linewidth]{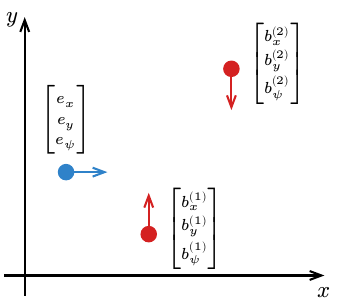}
    \caption{Ego vehicle $e$ and two obstacles $\obs^{(1)}, \obs^{(2)} \in \Obs$ positioned in the global coordinate frame.}
    \label{figApplicationEgoAndObstacleInGlobalReferenceFrame}
\end{figure}

With these definitions the ego vehicle, together with the moving obstacles, can now be placed in the global coordinate frame, i.e., the 2-dimensional Euclidean space, depicted in \cref{figApplicationEgoAndObstacleInGlobalReferenceFrame}.

\subsection{Interpretation}\label{secApplicationPrerequisitesInterpretation}

In order to describe which areas are likely to be occupied by an obstacle, we model the obstacle's location $\vecrow{x, y} \in \bb{R}^2$ probabilistically in the space $\bb{R}^2$, conditioned on time $t \in T$. Thus, the occupancy distribution for an obstacle $\obs \in \Obs$ at time $t$ is
\begin{equation*}
  p_\obs (x,y | t), 
\end{equation*}
such that
\begin{equation*}
  \iint_{\bb{R}^2} p_\obs (x,y | t) \dif x \dif y = 1.
\end{equation*}

These distributions could be generated by uncertainty propagation methods. For this work we assume them to be provided.

We further assume these occupancy distributions to be freely moved and rotated to align with the obstacle's state $\obs(t_0)$ at the beginning $t_0$ of the time interval $T$. This implies the dynamics of the obstacles to be invariant with regards to transformations of the initial state $\obs(t_0)$. Please consult \cite[][ch. 2 and appx. A]{murray1994} for a detailed discussion. It is safe to make this assumption for all the ground based obstacles that we like to consider.

By this property, each occupancy distribution $p_\obs$ can be defined for a nominal state $\obs^{(n)} = 0$. I.e., for a vehicle starting at the origin, and heading in the direction of the $x$-axis. 
Let $\phi_\obs$ be a function that maps the location of obstacle $\obs$'s initial state $\obs(t_0)$ to the nominal state $\obs^{(n)}$.

It is $R(\psi) \in \bb{R}^{2 \times 2}$ a 2-dimensional rotation matrix for an angle of $\psi$ such that
\begin{align*}
      \phi_\obs &: \bb{R}^3 \to \bb{R}^2, \\
          & \begin{bmatrix}
            \obs_x (t) \\
            \obs_y (t) \\
            \obs_\psi (t)
        \end{bmatrix} \mapsto R(-\obs_\psi (t_0)) \left(\begin{bmatrix}
            \obs_x (t) \\
            \obs_y (t)
        \end{bmatrix} - \begin{bmatrix}
            \obs_x (t_0) \\
            \obs_y (t_0)
        \end{bmatrix}\right).
\end{align*}

With this we express the probability distribution $p_\obs (x, y | t)$ of the obstacle $\obs \in \Obs$ in terms of the corresponding nominal distribution $p_n (x, y | t)$ by applying $\phi_\obs$
\begin{equation}
  p_\obs (\obs_x (t), \obs_y (t) | t) \coloneqq p_n (\phi_\obs (\obs (t)) | t).
  \label{eqApplicationOccupancyDistributionComposeStateTransform}
\end{equation}

Besides the smoothness requirement our definition of risk is required to answer the question of how to balance the simplifying assumption of point masses for the ego vehicle and the obstacles, and how to combine the occupancy distributions of multiple obstacles in the environment.

In regard to the former, it is desirable to not only consider the location of the ego vehicle. To simplify the notation we introduce the trajectory $\gamma$ of the ego vehicle $e$ as
\begin{equation}
    \begin{aligned}
      \gamma: T &\to \bb{R}^2,\\
        t &\mapsto \begin{bmatrix}
            e_x(t) \\
            e_y(t)
        \end{bmatrix}
    \end{aligned}
     \label{eqApplicationTrajectoryOfEgoVehicle}
\end{equation}

the projection of the ego vehicle's state $e(t)$ at time $t \in T$ onto its $x$-$y$ components. Instead of the points on the trajectory we consider a disk $\cDisk(\gamma(t)) \subset \bb{R}^2$ around the trajectory at time $t$. The actual size of this disk could be derived from the ego vehicle's and the obstacle's physical dimensions (e.g., half-lengths of ego vehicle + the obstacle + extra safety buffer). By using the disk we ask, "is the ego vehicle in a risky region considering its physical footprint" instead of "is the ego vehicle at a risky point".

We now consider the challenge of merging multiple occupancy distributions. A direct summation of likelihoods may yield probabilities greater than one, thus losing the probabilistic interpretation. Normalizing by the number of obstacles, on the other hand, can lead to implausibly low collision probabilities or require disjoint supports to circumvent such normalization. To address these issues, we investigate three distinct interpretations.

Instead of directly adding probabilities of a collision, we could interpret the problem as an event of "avoiding all obstacles" (cf. \cite{dutoit2011}). 
The probability of collision-free traversal, i.e., staying safe along a trajectory $\gamma$, is the product of the complements of the collision probabilities with each obstacle:
\begin{equation*}
  P_{\text{safe}} (\gamma) = \prod_{\obs \in \Obs} \left(1 - \int_{t \in T} \iint_{\cDisk(\gamma(t))} p_\obs (x,y | t) \dif x \dif y \dif t\right).
\end{equation*}

Hence, the probability of a collision is
\begin{equation*}
  P_{\text{collision}} (\gamma) = 1 - P_{\text{safe}} (\gamma).
\end{equation*}

This always yields a well-defined value within the interval $\interval{0}{1}$, ensuring a probabilistic interpretation. 
The major drawback of this view is numerically. If collision probabilities are high or trajectories are long, the product can become very small.
Something that could be fixed by working with the $\log$-likelihoods and adopting a interpretation of collision entropy or collision information.

A second perspective frames risk using Conditional Value-at-Risk (CVaR), as discussed in works such as \cite{hakobyan2019, majumdar2020}. With $\alpha \in \left(0,1\right)$, this measure captures the expected risk of outcomes within the worst $(1-\alpha) \cdot 100\%$ of collision scenarios. The associated Value-at-Risk (VaR) determines the greatest risk still within the least severe $\alpha \cdot 100\%$ of cases. CVaR is then defined as the expected risk beyond this point. Tuning $\alpha$ adjusts the sensitivity to rare but high-risk events in the tail of the distribution.

Formally, this can be defined by introducing a binary random variable $C_\obs (\gamma)$ indicating collisions (1 if trajectory $\gamma$ collides with $\obs \in \Obs$, 0 otherwise). The probability for a collision, i.e., $C_\obs (\gamma) = 1$, is given by the integral 
\begin{equation*}
  \frac{1}{\abs{T}} \int_{t \in T} \iint_{\cDisk(\gamma(t))} p_\obs (x,y | t) \dif x \dif y \dif t.
\end{equation*}
The total number of collision is therefore $C(\gamma) = \sum_{\obs \in \Obs} C_\obs (\gamma)$. Given the inverse cumulative distribution function (iCDF) of $C(\gamma)$, the VaR for $\alpha$ is 
\begin{equation*}
  \nu = F^{-1}_{C(\gamma)} (\alpha).
\end{equation*}
The respective CVaR is therefore
\begin{equation*}
  \text{CVaR}_\alpha (\gamma) = \bb{E} \left[C | C > \nu\right].
\end{equation*}

CVaR explicitly addresses the heavy-tail phenomenon common in practical risk assessments and is therefore indispensable when the objective is to prevent severe outcomes. Under this interpretation, we ask: "Given that we accept an $\alpha \cdot 100\%$ risk level, what is the expected number of collisions in the worst $(1-\alpha) \cdot 100\%$ of cases?". However, computing CVaR increases complexity, as it typically requires sampling-based or approximation techniques, which may render the approach computationally infeasible in real-time scenarios.

Similar to approaches of \cite{dutta2020, tipaldi2011}, we consider a further perspective in which obstacle occurrences are modeled as events in a spatio-temporal Poisson random field with intensity $\lambda : \bb{R}^2 \times T \to \left[0, \oo\right)$, 
\begin{equation}
  \lambda (x, y; t) = \sum_{\obs \in \Obs} p_\obs (x,y | t).
  \label{eqApplicationTowardsProblemOccupancyDistributionAndRiskIntensityFunction}
\end{equation}

In this formulation, the occupancy distribution is reinterpreted as a spatial occurrence rate. As a result, aggregating likelihoods becomes meaningful without the need to artificially constrain the values to the interval $\interval{0}{1}$. Integrating the resulting intensity over the spatio-temporal region defined by the trajectory $\gamma$ and its associated disks $\cDisk(\gamma)$ yields the expected number of collisions:
\begin{equation*}
  \Lambda(\gamma) = \int_{t \in T} \iint_{\cDisk(\gamma (t))} \lambda (x,y; t) \dif x \dif y \dif t.
\end{equation*}

In a generic Poisson random field, the integral $\Lambda(\gamma)$ may diverge or attain arbitrarily large values. This is not the case in our setting: at each instant $t \in T$, each obstacle must occupy some location in space, implying that the spacial integral of the intensity equals the number of obstacles $\abs{\Obs}$. Consequently, when integrating over the time interval $T$ of length $\abs{T}$, we obtain:
\begin{equation}
  \int_{t \in T} \iint_{\bb{R}^2} \lambda (x,y; t) \dif x \dif y \dif t 
  = \int_{t \in T} \abs{\Obs} \dif t 
  = \abs{\Obs} \abs{T}.
  \label{eqApplicationIntensityFunctionBoundedness}
\end{equation}

This represents the total number of obstacle occurrences over the entire spatio-temporal domain, including regions further away from the trajectory. Defining acceptable thresholds for collision risk based on this formulation may depend on empirical data and safety criteria.

Defining risk as the probability that a trajectory remains collision-free offers the advantage of being a direct probabilistic interpretation. However, this approach is only suitable for sufficiently large probabilities in order to avoid numerical issues or the need for more abstract interpretations.

The CVaR-based perspective is particularly valuable when the tail behavior of the distribution of the number of collisions plays a dominant role. However, this is not relevant in our scenario, where the primary objective is still to avoid collisions, and risk acceptance is only considered when dictated by external circumstances.

This leads us to favor the intensity-based interpretation using a Poisson random field. This is appropriate, as the accumulated occupancy distributions can be naturally reinterpreted as spatio-temporal collision rate. Moreover, the assumptions required to define a Poisson random field are satisfied:
\begin{itemize}
    \item The total intensity is finite (see \cref{eqApplicationIntensityFunctionBoundedness}).
    \item For each $t \in T$, the disk $\cDisk(\gamma(t))$ is commonly an open or closed subset of $\bb{R}^2$, and therefore Borel measurable (cf. \cite{meintrup2005}).
    \item Independency of point occurrences is satisfied, as no interaction among obstacles is assumed.
\end{itemize}

\subsection{B-Spline Surface}\label{secApplicationPrerequisitesBSplineSurface}

To ensure the differentiability required by certain optimization methods, such as Newton-type methods or sequential quadratic programming, we require the Poisson intensity $\lambda (x, y; t)$ to lie within the Sobolev space $W_{d,\oo}$. This space is defined as
\begin{align*}
  W_{d,\oo} &= W_{d,\oo} (\bb{R}^2 \times T, \bb{R}) \\
    &= \left\{f : \bb{R}^2 \times T \to \bb{R} : \norm{f}_{L_\oo} < \oo \text{ and } \norm{D^i f}_{L_\oo} < \oo, \forall i \in \left\{1, \ldots, d\right\}\right\},
\end{align*}
and contains all the functions from $\bb{R}^2 \times T$ to $\bb{R}$ that are $d$-times differentiable and whose derivatives up to order $d$ ($D f, ..., D^d f$) have bound Lebesgue norms ($L_\oo$-norm). Membership in $W_{d,\oo}$ guarantees the existence and boundedness of gradients and higher-order derivatives, and promotes numerical stability when used in conjunction with derivative-based OCP solvers.

Depending on the chosen uncertainty propagation method, occupancy distributions may be represented either as continuous probability density functions (PDF) or discrete probability mass functions (PMF). Nighter representation inherently satisfies the structural requirements of the Sobolev space.

To ensure differentiability, we approximate each occupancy distributions $p_\obs$ by a smooth B-spline surfaces $\tilde{p}_\obs$. B-splines of degree $d$ (i.e., order $d+1$) span a finite-dimensional subspace of $W_{d,\oo}$, thereby ensuring existence and boundedness of derivatives. B-splines offer several advantages in this context:
\begin{itemize}
    \item The smoothness is directly controlled by the spline degree $d$, 
    \item B-splines and their derivatives are computationally efficient to construct and evaluate, 
    \item approximation quality and computational cost can be balanced by adjusting the number and placement of the spline knots, and
    \item due to their compact support, the support of the approximation $\tilde{p}_\obs$ closely matches that of the original occupancy distribution $p_\obs$, avoiding artificial spreading.
\end{itemize}

Because the intensity $\lambda$ is a linear combination of B-spline approximations $\tilde{p}_\obs$, it follows that $\lambda \in W_{d,\oo}$. Note that this holds also for the composition $p_n \circ \phi_\obs$ (see \cref{eqApplicationOccupancyDistributionComposeStateTransform}).

To prepare for the regularity analysis, we introduce B-splines appropriate to our use case. The goal is to approximate the occupancy distributions $p (x, y | t)$, a mapping from $\bb{R}^2 \times T$ to $\interval{0}{1}$, using non-negative bivariate B-Spline surfaces. These take the form:
\begin{equation*}
  \hat{p} (x,y; t) = \sum_{i=1}^{n_x+d_x-1} \sum_{j=1}^{n_y+d_y-1} \hat{c}_{i,j} (t) B_i^{d_x} (x) B_j^{d_y} (y),
\end{equation*}
where the spline coefficients are non-negative (i.e., $\hat{c}_{i,j} (t) \ge 0$). These coefficients are collected in the spline coefficient tensor 
\begin{equation*}
  \hat{C} = \begin{bmatrix}
      \hat{c}_{i,j}(t)
  \end{bmatrix}_{\substack{
      i = 1, ..., n_x+d_x-1\\ 
      j = 1, ..., n_y+d_y-1
  }}.
\end{equation*}
The univariate B-spline basis function $B_i^{d_x} (x)$ and $B_j^{d_y} (y)$ with $i = 1, \ldots, n_x+d_x-1$ and $j = 1, \ldots, n_y+d_y-1$ are non-negative, and of degree $d_x$ and $d_y$, respectively. These functions are constructed according to the de Boor scheme (cf. \cite{deboor1977, piegl1996}).

To ensures a consistent approximation structure across time, we employ a uniform knot placement strategy over the spatial domain 
\begin{equation*}
  \interval{x_{\min}}{x_{\max}} \times \interval{y_{\min}}{y_{\max}} \subset \bb{R}^2
\end{equation*}
that fully encompasses the supports of the occupancy distribution $p$ over time. 

To guarantee a smooth decay to zero outside the approximation domain we add knots of multiplicity $d_x+1$ or $d_y+1$ at both ends in the $x$- and $y$-direction, respectively. Choosing the number of uniformly spaced interior points $n_x$ and $n_y$, the total number of knots becomes $n_x + 2 d_x$ in the $x$-direction and $n_y + 2 d_y$ in the $y$-direction. Thus, the resulting knots in the $x$-direction are defined as
\begin{equation*}
  \xi_i = \begin{cases}
    x_{\min} &\text{if } 1 \le i \le d_x + 1 \\
    x_{\min} + (i - d_x - 1) \frac{x_{\max} - x_{\min}}{n_x - 1} &\text{if } d_x + 2 \le i \le n_x + d_x - 1 \\
    x_{\max} &\text{if } n_x + d_x \le i \le n_x + 2 d_x
  \end{cases}.
\end{equation*}

Similarly, the knots in the $y$-direction are given by:
\begin{equation*}
  \nu_j = \begin{cases}
    y_{\min} &\text{if } 1 \le j \le d_y + 1 \\
    y_{\min} + (j - d_y - 1) \frac{y_{\max} - y_{\min}}{n_y - 1} &\text{if } d_y + 2 \le j \le n_y + d_y - 1 \\
    y_{\max} &\text{if } n_y + d_y \le j \le n_y + 2 d_y
  \end{cases}.
\end{equation*}

It is important to note that this uniform knot placement strategy can be inefficient when the occupancy distribution exhibits significant variation or contains fine structural features. In such cases, adaptive knot placement could improve approximation accuracy. However, adaptive schemes conflict with our intend of a consistent approximation structure, which helps to smoothly interpolate spline coefficients $\hat{c}_{i, j} (t)$ over time. This interpolation is necessary, as occupancy distributions might only be provided for discrete time steps.

To preserve the interpretation of $\hat{p}$ as a probability distribution, it is necessary to normalize the B-spline approximation. Exploiting the tensor-product structure of the B-spline surface, the integration of the surface $\hat{p}$ over the domain becomes straightforward:
\begin{align*}
  V(\hat{p}; t) &= \iint_{\bb{R}^2} \hat{p}(x,y; t) \dif x \dif y \\
    &= \iint_{\bb{R}^2} \sum_{i=1}^{n_x+d_x-1} \sum_{j=1}^{n_y+d_y+1} \hat{c}_{i,j} (t) B_i^{d_x} (x) B_j^{d_y} (y) \dif x \dif y \\
    &= \sum_{i=1}^{n_x+d_x-1} \sum_{j=1}^{n_y+d_y+1} \hat{c}_{i,j} (t) \left(\int_{x_{\min}}^{x_{\max}} B_i^{d_x} (x) \dif x\right) \left(\int_{y_{\min}}^{y_{\max}} B_j^{d_y} (y) \dif y\right) \\
    &= \sum_{i=1}^{n_x+d_x-1} \sum_{j=1}^{n_y+d_y+1} \hat{c}_{i,j} (t) I_x (i) I_y (j).
\end{align*}

Here, the total volume $V$ is computed as a weighted sum of the products of the univariate B-spline integrals:
\begin{align*}
  I_x (i) &= \int_{x_{\min}}^{x_{\max}} B_i^{d_x} (x) \dif x &&= \int_{\xi_i}^{\xi_{i+d_x+1}} B_i^{d_x} (x) \dif x, \quad i = 1, \ldots, n_x + d_x - 1, \\
  I_y (j) &= \int_{y_{\min}}^{y_{\max}} B_j^{d_y} (y) \dif y &&= \int_{\nu_j}^{\nu_{j+d_y+1}} B_j^{d_y} (y) \dif y, \quad j = 1, \ldots, n_y + d_y - 1.
\end{align*}
These integrals can be precomputed and stored for efficient reuse. 

Finally, normalization is accomplished by dividing the entire surface $\hat{p}$ by the total volume $V(\hat{p}; t)$, resulting in scaled coefficients $c_{i,j} (t) \coloneqq \frac{\hat{c}_{i,j} (t)}{V(\hat{p}; t)}$: 
\begin{equation}
    \begin{aligned}
      \tilde{p} (x, y; t) &:= \frac{1}{V(\hat{p}; t)} \hat{p}(x, y; t) \\
        &= \frac{1}{V(\hat{p}; t)} \sum_{i=1}^{n_x+d_x-1} \sum_{j=1}^{n_y+d_y+1} \hat{c}_{i,j} (t) B_i^{d_x} (x)  B_j^{d_y} (y) (y) \\
        &= \sum_{i=1}^{n_x+d_x-1}\sum_{j=1}^{n_y+d_y+1} \underbrace{\frac{\hat{c}_{i,j} (t)}{V(\hat{p}; t)}}_{ c_{i,j} (t)} B_i^{d_x} (x)  B_j^{d_y} (y).
    \end{aligned}
    \label{eqApplicationNormalizationOfBSplineSurface}
\end{equation}

\section{Approximation Problem}\label{secApplicationApproximationProblem}

Approximating non-smooth functions using smooth basis functions introduce several challenges (cf. \cite{wahba1990}). To illustrate this, consider the example shown in \cref{imgApplicationAnalysisTC3OPDAndIntensityMap-original}, which depicts the nominal occupancy distribution $p_n (x, y | t=3.5s)$ for a small car modeled with single-track kinematics. 

The plot reveals regions of nearly constant likelihoods (plateaus), sharp discontinuities (e.g., the tightly packed blue and green colored contour lines), and narrow corridors such as the region around the $x$-axis from $x \approx 25$ to $x \approx 35$. These features pose difficulties for smooth approximations: plateaus may induce spikes, sharp transitions may cause oscillations, and fine structures may be over-smoothed. This observation motivates the central challenge addressed in this work.

Our goal is to find an approximation $\tilde{p}$ of the occupancy distribution $p$ that preserves as much of the features as possible while being sufficiently smooth to facilitate the generation of a risk-minimizing trajectory. To this end, we introduce three metrics that form the basis of the objective function used to calculate the B-spline coefficients. 

The first metric quantifies how closely $\tilde{p}$ approximates the occupancy distribution $p$ over the spatial domain $\bb{R}^2$. We use the mean square error (MSE) to measure fidelity:
\begin{equation}
    \begin{aligned}
        \opMSE(\tilde{p}; t) = &\norm{p (x, y | t) - \tilde{p}(x, y | t)}^2 \\
        = &\iint_{\bb{R}^2} \left((p (x, y | t) - \tilde{p} (x, y | t))\right)^2 \dif x \dif y\\
        = &\iint_{\bb{R}^2} \left(p (x,y|t) - \sum_{i=1}^{n_x+d_x-1}\sum_{j=1}^{n_y+d_y+1} c_{i,j} (t) B_i^{d_x} (x)  B_j^{d_y} (y) \right)^2 \dif x \dif y.
    \end{aligned}
    \label{eqApplicationProblemSmoothingMeanSquareError}
\end{equation}

To suppress oscillations or spikes, we incorporate regularization terms (cf. \cite{unser1993, eilers1996}). Three common strategies include:
\begin{itemize}
    \item adaptive knot placement,
    \item penalization of the total variation, and
    \item penalization of higher-order derivatives.
\end{itemize}
While adaptive knot placement can improve local fidelity and regularity of the approximation, we omit it here due to the previously mentioned requirement of a consistent structure and focus instead on derivative-based regularization.

The first regularization metric is the total variation (TV) of $\tilde{p}$, which penalizes large gradients and favors piecewise constant approximations (cf. \cite{rudin1992}). This helps suppress oscillations and spikes while preserving sharp transitions (edges).

The gradient of $\tilde{p}$ at time $t$ is defined as:
\begin{equation*}
  \nabla \tilde{p} (x, y | t) = \begin{bmatrix}
      \frac{\partial\tilde{p}}{\partial x} \\
      \frac{\partial\tilde{p}}{\partial y}
  \end{bmatrix}.
\end{equation*}

The continuous form of total variation is given by the integral of the squared gradient norm over the spacial domain:
\begin{equation}
    \begin{aligned}
        \opTV(\tilde{p}; t) &= \iint_{\bb{R}^2} \norm{\nabla \tilde{p}}^2 \dif x \dif y \\
    &= \iint_{\bb{R}^2} \left(\left(\frac{\partial \tilde{p}}{\partial x}\right)^2 + \left(\frac{\partial \tilde{p}}{\partial y}\right)^2\right) \dif x \dif y
    \end{aligned}
    \label{eqApplicationTotalVariationContinuos}
\end{equation}

Note that this formulation deviates from the classical definition of total variation, which does not square the gradient norm. This modification is intentional, as it allows for a smooth differentiable regularization term, which is required in the subsequent approximation formulated as an minimization problem.

To simplify the calculation, we adopt the approximation strategy proposed by \cite{eilers1996} and replace continuous derivatives with finite differences of adjacent B-spline coefficients:
\begin{equation*}
  \opTV_\text{approx}(\tilde{p}; t) = \sum_{i=1}^{n_x+d_x-2} \sum_{j=1}^{n_y+d_y-2} \left(c_{i+1,j} (t) - c_{i,j} (t)\right)^2 + \left(c_{i,j+1} (t) - c_{i,j} (t)\right)^2.
\end{equation*}

As a second regularization metric we penalize curvature using Tikhonov regularization (cf. \cite{wahba1990}). This discourages large second derivatives and as a effect promotes global smoothness and suppresses oscillatory behavior. 

The continuous form is defined as:
\begin{equation}
    \opT(\tilde{p}; t) = \iint_{\bb{R}^2} \left(
        \left(\frac{\partial^2\tilde{p}}{\partial x^2}\right)^2 
        + \left(\frac{\partial^2\tilde{p}}{\partial x \partial y}\right)^2 
        + \left(\frac{\partial^2\tilde{p}}{\partial y^2}\right)^2
    \right) \dif x \dif y.
    \label{eqApplicationTikhonovContinuos}
\end{equation}

Following \cite{eilers1996}, we approximate this with finite differences of neighboring spline coefficients (the time argument of the coefficients are omitted for brevity):
\begin{equation*}
    \begin{aligned}
        \opT_\text{approx} (\tilde{p}; t) = &\sum_{i=2}^{n_x+d_x-2} \sum_{j=2}^{n_y+d_y-2} \left(
            \left(c_{i+1,j} - 2\,c_{i,j} + c_{i-1,j}\right)^2 \right.\\
            &+ 2\,\left(c_{i+1,j+1}-c_{i+1,j-1}-c_{i-1,j+1}+c_{i-1,j-1}\right)^2 \\
            &\left.+ \left(c_{i,j+1} - 2\,c_{i,j} + c_{i,j-1}\right)^2
        \right).
    \end{aligned}
\end{equation*}

Using these metrics, we formulate the following minimization problem to compute the B-spline coefficients: 
\begin{equation*}
    \begin{aligned}
  \underset{C}{\min} &\quad \rhoMSE \cdot \opMSE(\tilde{p}; t) + \rhoTV \cdot \opTV(\tilde{p}; t) + \rhoT \cdot \opT(\tilde{p}; t) \\
  \text{s.t.} &\quad c_{i,j} (t) \ge 0, \quad i = 0, \ldots, n_x + d_x - 1,\quad j = 0, \ldots, n_y + d_y - 1
    \end{aligned}
    \label{eqApplicationApproximationProblem}
\end{equation*}

By adjusting the weights $\rhoMSE$, $\rhoTV$ and $\rhoT$ we control the trade-off between fidelity, variation, and smoothness in the final approximation $\tilde{p}$. 

Two additional constraints could be considered:
\begin{enumerate}
    \item Enforcing $\tilde{p} (x, y | t) \ge p (x, y | t)$ for all $\vecrow{x, y} \in \bb{R}^2$ and $t \in T$, to ensure conservative estimates. However, this may result in overly cautious behavior and suboptimal trajectories.
    \item Requiring normalization $\iint_{\bb{R}^2} \tilde{p} (x, y; t) \dif x \dif y = 1$. This was omitted due to inconsistent results observed with the used optimization solver. Instead, we apply normalization as a post-processing step (see \cref{eqApplicationNormalizationOfBSplineSurface}).
\end{enumerate}

The approximation $\tilde{p}$ we obtained under consideration of the regularization from the occupancy distribution will be called intensity map, in reference to the intensity function $\lambda$.

\section{Evaluation} \label{secApplicationEvaluation}

In this section, we define the metrics and procedures used to evaluate how different regularization methods -- Total Variation (TV) and Tikhonov -- affect approximation fidelity, variability, smoothness, and downstream OCP solver performance.

First, we empirically benchmark multiple parameter sets, i.e., combinations of regularization weights $\rhoMSE$, $\rhoTV$ and $\rhoT$, on a suite of test cases (TCs) that represent occupancy distributions with varying stochastic structures and complexity. From the results, we select three parameter sets: one that minimizes MSE, one that minimizes TV, and one that achieves a balance between MSE and TV.

In the second phase, these parameter sets are applied to a time-series of occupancy distributions representing the motion of a vehicle. The obtained approximations are integrated into an OCP to solve a representative path planning scenario. By solving this OCP repeatedly under slight variations of the initial conditions, we obtain statistical insights into how the chosen regularization influences the OCP solver's performance.

\subsection{Metrics} \label{secApplicationEvaluationMetrics}
We evaluate the effects of the regularization using two distinct sets of metrics.

The first set corresponds to metrics introduced in \cref{secApplicationApproximationProblem} and are used to assess the fidelity, variation, and smoothness of the approximation:
\begin{enumerate}
    \item MSE (\cref{eqApplicationProblemSmoothingMeanSquareError}): quantifies how closely the approximation $\tilde{p}$ follows the original occupancy distribution $p$,
    \item Supremum of the gradient's norm $\norm{\nabla \tilde{p}}$: helps detect potential over-smoothing by indicating loss of edge strength, 
    \item Supremum of the second derivative (cf. the integrand in \cref{eqApplicationTikhonovContinuos}): identifies regions of high curvature, which could signal artifacts like spikes,
    \item TV (\cref{eqApplicationTotalVariationContinuos}): measures the overall surface variability. 
\end{enumerate}

The second set of metrics captures the impact of regularization on the OCP solver performance:
\begin{enumerate}
    \item number of solver iterations,
    \item convergency time, and
    \item final value of the OCP objective function.
\end{enumerate}

For a summary of all metrics used in the evaluation, see \cref{tblApplicationEvaluatingSmoothingEvaluationMetrics}.

\begin{table}
    \centering
    \begin{tabularx}{.95\textwidth}{cXX}
        \toprule
        \multicolumn{2}{l}{\textbf{Metric}} & \textbf{Purpose}\\
        \midrule
        M1 & Mean Square Error (MSE) & Quantify approximation error\\
        M2 & Supremum of $\tilde{p}$'s gradient norm & Detect potential over-smoothing\\
        M3 & Supremum of $\tilde{p}$'s second derivative & Detect the presence of regions with high curvature\\
        M4 & Total Variation (TV) of $\tilde{p}$ & Measure overall surface variability\\
        \midrule
        M5 & Number of OCP solver iterations & How hard it is to optimize over the risk field\\
        M6 & Time for the OCP solver to converge & Efficiency for MPC loop\\
        M7 & Final objective function value in the OCP & Practical trajectory quality\\
        \bottomrule
    \end{tabularx}
    \caption{Metrics to quantify the quality of the approximation and the effect on the OCP solver performance.}
    \label{tblApplicationEvaluatingSmoothingEvaluationMetrics}
\end{table}

\subsection{Test Cases for the Approximation of Occupancy Distributions} \label{secApplicationEvaluationApproximationOfOccupancyDistribution}
We define three probability distributions of varying stochastic structure and complexity to serve as our test cases (TCs). These cases are used to evaluate different regularization configurations, i.e., the chosen weights $\rhoMSE$, $\rhoTV$ and $\rhoT$. An overview of the TCs is presented in \cref{tblApplicationEvaluatingSmoothingTestCases}.

\begin{table}
    \centering
    \begin{tabularx}{.95\textwidth}{cXX}
        \toprule
        \textbf{Test Case} & \textbf{Description} & \textbf{Structure}\\
        \midrule
        TC1 & Product of two independent Gaussian distributions. & smooth and independent\\
        TC2 & Gaussian in the $y$ with variance conditioned on a Beta distribution in the $x$. & smooth and dependent\\
        TC3 & Occupancy distribution from a uncertainty propagation method. & rough, plateaus, fine structures, and dependent\\
        \bottomrule
    \end{tabularx}
    \caption{Test Cases used for evaluating the approximation of different regularization configurations.}
    \label{tblApplicationEvaluatingSmoothingTestCases}
\end{table}

Let $\vecrow{X, Y}$ be a pair of random variables. In TC1 this pair follows a bivariate normal distribution with mean $\mu$ and diagonal covariance matrix $\Sigma$, with the variances $\sigma_x^2$ and $\sigma_y^2$ in the respective dimensions and no correlation:
\[
  \mu = \begin{bmatrix}
      0 \\
      0
  \end{bmatrix}, \quad \Sigma = \begin{bmatrix}
      \sigma_x^2 & 0 \\
      0 & \sigma_y^2
  \end{bmatrix}.
\]
The joint PDF is:
\[
  p_1(x, y) = \frac{1}{2 \pi \sigma_x \sigma_y} \exp\left(-\frac{1}{2} \left(\left(\frac{x}{\sigma_x}\right)^2+\left(\frac{y}{\sigma_y}\right)^2\right)\right).
\]

TC2 defines a joint distribution where:
\begin{itemize}
    \item The marginal distribution of $X$ is a Beta distribution $f_X (x; \alpha, \beta)$,
    \item The conditional distribution of $Y$ given $X = x$ is a truncated normal distribution with mean zero and variance $\sigma^2(x)$, which depends on the likelihood of $f_X$.
\end{itemize}

The respective PDFs are:
\begin{align*}
    f_X (x; \alpha, \beta) &= \frac{x^{\alpha-1} (1 - x)^{\beta-1}} {B(\alpha, \beta)}, &0 \le x \le 1,\\
    f_{Y|X} (y; \sigma^2(x)) &= \frac{1}{\sqrt{2 \pi \sigma^2(x)}} \exp\left(-\frac{y^2}{2 \sigma^2(x)} \right), &y \in \interval{-h}{+h},
\end{align*}
where $B(\alpha, \beta)$ is the Beta function and $\sigma(x)$ is a non-negative function given by:
\[
  \sigma(x) = 0.1 + \frac{f_X (x; \alpha, \beta)}{\inf_{u \in \interval{0}{1}} f_X (u; \alpha, \beta)}.
\]

This definition causes the spread of the conditional normal distribution to vary smoothly with the Beta distributions likelihood. The resulting joint PDF is:
\[
  p_2 (x,y; \alpha, \beta, \sigma) = f_X (x; \alpha, \beta) \cdot f_{Y|X} (y; \sigma^2(x))
\]
with support $\interval{0}{1} \times \interval{-h}{+h}$.

A contour plot of this occupancy distribution is shown in \cref{imgApplicationAnalysisTC2OdVsIntensityMap-original}.

TC3 uses an occupancy distribution derived from an uncertainty propagation method applied to a vehicle modeled with a single-track kinematic model. This propagation approach leverages the structure of the system's state trajectories, in contrast to conventional methods that propagate uncertainty via transformations of sets or probability distributions through system dynamics. See \cref{imgApplicationAnalysisTC3OPDAndIntensityMap-original} for an example of distribution generated by this method.

The core idea is to represent uncertainty as qualitative expert knowledge, encoded in the form of an OCP. The state space is first partitioned into discrete cells, and a series of OCPs is solved to reach each individual cell. Each resulting trajectory is treated as a representative sample for all trajectories terminating in its vicinity. These trajectories are then quantized into motion primitives, which serve as building blocks. By evaluating the likelihood of individual sequences of motion primitives, the probability of occupancy for each corresponding cell can be estimated. A full description of this method is currently in preparation for publication.

Note: For TC1 and TC2, the constructed distributions $p_i$, $i \in {1, 2}$ are static and do not include a time dependence (i.e., the conditioning on the variable $t$ in $p(x, y | t)$) and is therefore omitted for simplicity. TC3, however, reflects the dynamic nature of time-evolving occupancy distributions.

\subsection{Results for the Test Cases} 
Our evaluation of the approximation focuses on the influence of regularization rather than the impact of the knot configuration, such as the number of knots or the degree of the basis polynomials. We selected suitable configurations based on non-regularized B-spline approximations that most closely approximate the TCs. This selection was guided by minimizing the expansion of the support, measured relative to the support of the occupancy distribution, and the MSE. 

The number of interior points $n_x$ and $n_y$ evaluated in the $x$- and $y$-dimension respectively, are listed in \cref{tblApplicationKnotConfigurationNKnots}. These specify the associated knots for the univariate B-spline basis functions. Additionally, we considered both quadratic (degree 2) and cubic (degree 3) B-spline bases in each dimension.

\begin{table}
    \centering
    \begin{tabular}{lll}
        \toprule
        \textbf{} & \textbf{$n_x \in$} & \textbf{$n_y \in$}\\
        \midrule
        TC1 & $\left\{3, 5, 8, 10, 15, 20, 23, 25, 30\right\}$ & $\left\{3, 5, 8, 10, 15, 20, 23, 25, 30\right\}$\\
        TC2 & $\left\{4, 5, 10, 15, 20, 25, 30, 40\right\}$ & $\left\{4, 5, 10, 15, 20, 25, 30, 40\right\}$\\
        TC3 & $\left\{4, 5, 10, 15, 20, 25, 30, 39\right\}$ & $\left\{5, 10, 15, 16, 20, 25, 31, 46, 61\right\}$\\
        \bottomrule
    \end{tabular}
    \caption{Ranges of the number of knots for each TC to determine suitable knot configurations without regularization.}
    \label{tblApplicationKnotConfigurationNKnots}
\end{table}

The selected knot configurations, along with the degrees of the basis functions and the resulting approximation quality (in terms of MSE and support growth), are presented in \cref{tblApplicationKnotStrategyAndQualityWithoutRegularization}. These configurations serve as the fixed baseline for all subsequent regularization experiments.

\begin{table}
    \centering
    \begin{tabular}{l|llllll}
        \toprule
         & \multicolumn{4}{c}{\textbf{Knot Definition}} & \multicolumn{2}{c}{\textbf{Quality}} \\
         & $n_x$ & $d_x$ & $n_y$ & $d_y$ & \makecell{Relative change\\ in support area} & MSE\\
        \midrule
        TC1 & 25 & 3 & 15 & 3 & $\approx 0.77\%$ & $\approx 4.00\cdot10^{-10}$\\
        TC2 & 40 & 3 & 40 & 3 & $\approx 10\%$ & $\approx 1.13\cdot10^{-7}$\\
        TC3 & 20 & 3 & 25 & 3 & $\approx 2.4\%$ & $\approx 1.04\cdot10^{-8}$\\
        \bottomrule
    \end{tabular}
    \caption{Selected number of knots, degree of B-spline bases, and resulting approximation quality for each TC without regularization.}
    \label{tblApplicationKnotStrategyAndQualityWithoutRegularization}
\end{table}

Our approximation problem involves a multi-objective optimization setting in which three competing objectives -- MSE, TV, and Tikhonov regularization -- are combined. These objectives inherently conflict with one another: improving one may lead to a decline in another. In such a case, calculating a set of Pareto-optimal solutions illustrating the trade-off among the objectives is often more informative than a global optimum (cf. \cite{goodarzi2014}). 

We explore the Pareto front using the weighted sum method, with the weights $\rhoMSE$, $\rhoTV$ and $\rhoT$ varied systematically over the interval $\interval{0}{10}$. These weights represent the relative importance of each objective, and by sweeping across their values, we trace out the convex regions of the Pareto front. However, this approach may fail to capture non-convex portions of the front, which are better explored using alternative techniques such as the $\epsilon$-constraint method, normal boundary intersection (NBI), or evolutionary algorithms (cf. \cite{das1998, marler2010}). 

For the purpose of selecting suitable approximations, we focus on two objectives: M1 (MSE) and M4 (TV). The third metric, M3 (Tikhonov regularization), is excluded from the optimization process but still monitored.

For each TC, we present the evaluation results in the following format:
\begin{itemize}
    \item A plot of the Pareto front, showing the trade-off between MSE and TV,
    \item A visual comparison of the occupancy distribution, and the approximation as far as these are informative.
\end{itemize}

The plot of the Pareto front shows the approximation of the occupancy distribution, i.e., intensity maps, without any regularization as a red plus sign ($\rhoMSE = 10$, $\rhoTV = \rhoT = 0$). Blue x markers are used for weight configurations that have only the TV regularization $\rhoTV \neq 0$ and $\rhoT = 0$ (TV only), conversely, orange three-legged starfish marks show results of weight configurations of $\rhoTV = 0$ and $\rhoT \neq 0$ (Tikhonov only). Finally, green dots mark intensity maps obtained from weight configurations with both regularization terms active, i.e., $\rhoTV \neq 0$ and $\rhoT \neq 0$ (Combined). In any case, the weight $\rhoMSE$ is different from 0. 

\subsubsection{Test Case 1}
For TC1, the lowest MSE (M1) was achieved using no regularization, i.e., with both TV and Tikhonov weights set to 0. In contrast, the least variation (minimum TV, M4) was obtained with maximum weight on the TV regularization terms ($\rhoMSE = 0.5$, $\rhoTV = 9.5$, and $\rhoT = 0$).

The optimal trade-off between variability and accuracy was observed when using only TV regularization with a weight of 8.5, no Tikhonov regularization, and the weight 1.5 on MSE. Notably, the approximation error remained consistently low across all regularization configurations. This stability can be attributed to the inherent smoothness of the underlying occupancy distribution.

\begin{figure}
    \centering
    \begin{subfigure}{0.45\textwidth}
        \centering
        \includegraphics[width=\linewidth]{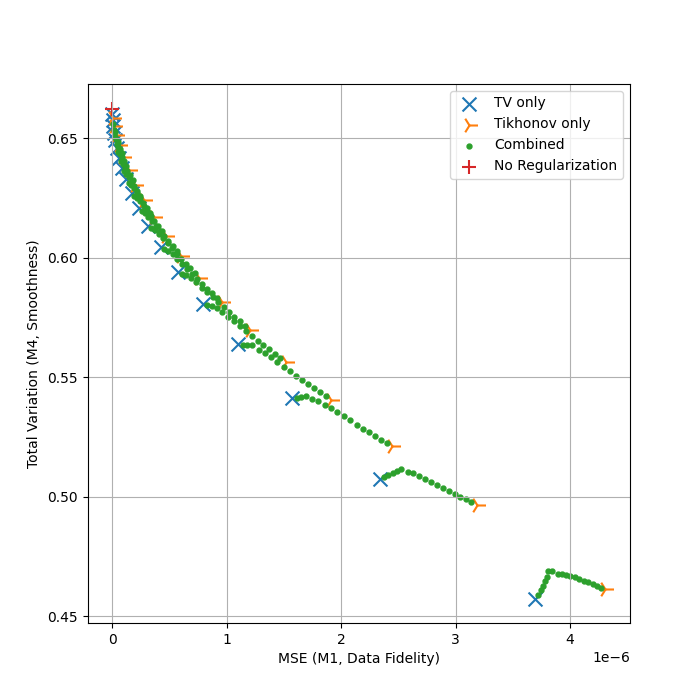}
        \caption{Test Case TC1.}
        \label{imgApplicationAnalysisOfEvaluationParetoFrontTC1}
    \end{subfigure}
    \hfill
    \begin{subfigure}{0.45\textwidth}
        \centering
        \includegraphics[width=\linewidth]{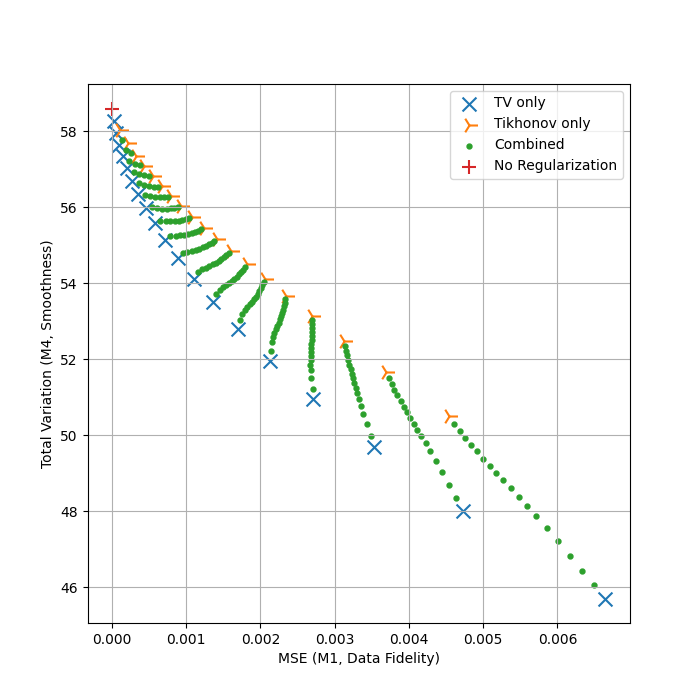}
        \caption{Test Case TC2.}
        \label{imgApplicationAnalysisOfEvaluationParetoFrontTC2}
    \end{subfigure}
    
    \vspace{1em}
    \begin{subfigure}{0.45\textwidth}
        \centering
        \includegraphics[width=\linewidth]{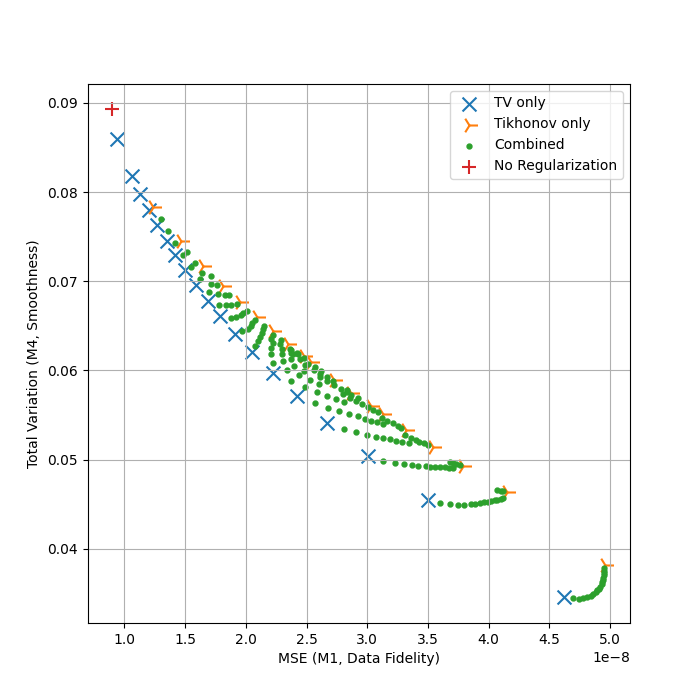}
        \caption{Test Case TC3.}
        \label{imgApplicationAnalysisOfEvaluationParetoFrontTC3}
    \end{subfigure}
    \hfill
    
    \caption{Pareto front illustrating the trade-off between fidelity (MSE) and variability (TV). Lower values are preferred.}
    \label{imgApplicationAnalysisOfEvaluationParetoFront}
\end{figure}

The resulting Pareto front, based on the tested weight configurations, is shown in \cref{imgApplicationAnalysisOfEvaluationParetoFrontTC1}.

\subsubsection{Test Case 2}
Although TC2 represents a smooth occupancy distribution, the average approximation error MSE (M1) of $2.1 \cdot 10^{-3}$ is approximately three orders of magnitude larger than that observed for TC1. The lowest MSE was achieved without any regularization (i.e., both regularization weights set to 0). Analogous to TC1, the lowest total variation (TV, M4) was obtained using the maximum weight on the TV regularization terms ($\rhoMSE = 0.5$, $\rhoTV = 9.5$, and $\rhoT = 0$). 

The best compromise between low MSE and TV occurred at using the weight configuration $\rhoMSE = 2$, $\rhoTV = 8$, and $\rhoT = 0$).

\begin{figure}
    \centering
    \begin{subfigure}{0.45\textwidth}
        \centering
        \includegraphics[width=\linewidth]{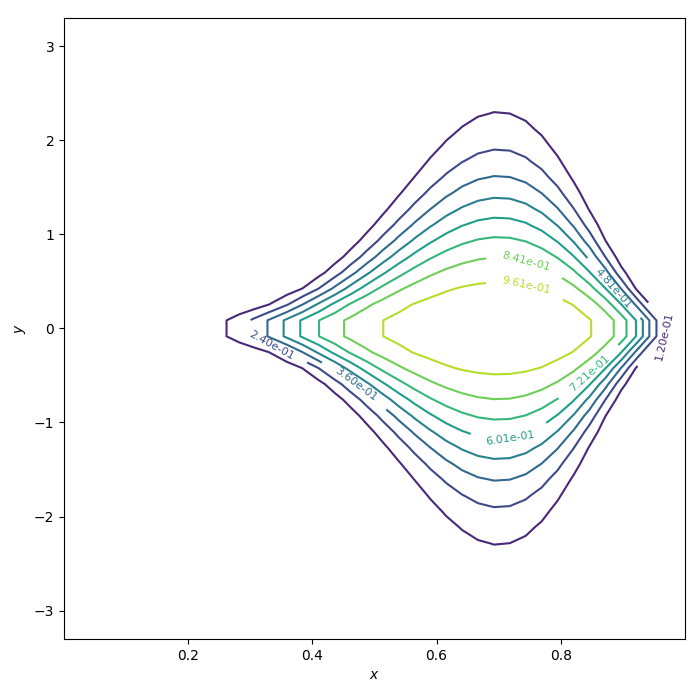}
        \caption{Occupancy distribution.\\\ }
        \label{imgApplicationAnalysisTC2OdVsIntensityMap-original}
    \end{subfigure}
    \hfill
    \begin{subfigure}{0.45\textwidth}
        \centering
        \includegraphics[width=\linewidth]{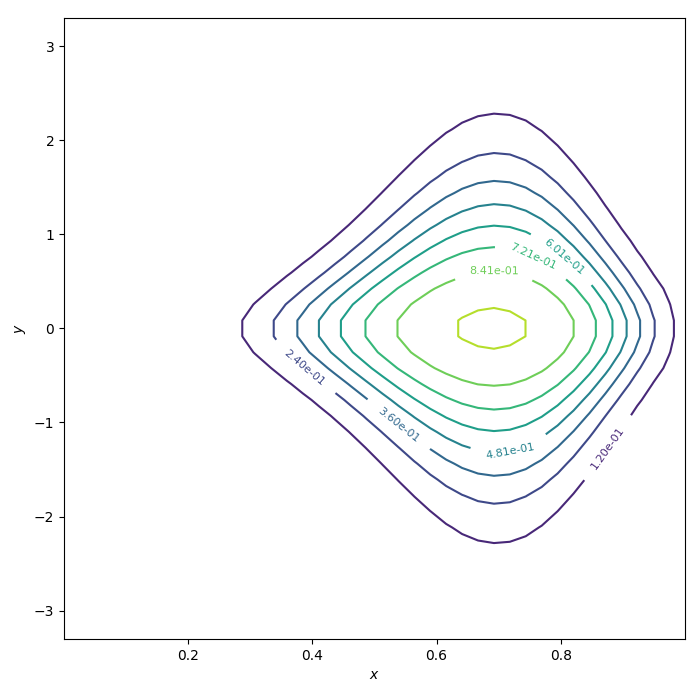}
        \caption{Intensity map with regularization weights $\rhoMSE = 2.0$, $\rhoTV = 8.0$, $\rhoT = 0$.}
        \label{imgApplicationAnalysisTC2OdVsIntensityMap-concavityStruggle}
    \end{subfigure}

    \caption{Comparison of occupancy distribution and intensity map for TC2.}
    \label{imgApplicationAnalysisTC2OdVsIntensityMap}
\end{figure}

A visual comparison between the occupancy distribution (\cref{imgApplicationAnalysisTC2OdVsIntensityMap-original}) and the corresponding approximation (\cref{imgApplicationAnalysisTC2OdVsIntensityMap-concavityStruggle}) reveals a notable struggle to capture the concavity of the distribution's support. This effect is more or less pronounced depending on the chosen regularization weights.

\Cref{imgApplicationAnalysisOfEvaluationParetoFrontTC2} shows the resulting Pareto front, based on the tested weight configurations.

\subsubsection{Test Case 3}
TC3 exhibits an overall low approximation error, with an average MSE (M1) of $2.8 \cdot 10^{-8}$, ranging from a minimum of $0.90 \cdot 10^{-8}$ to a maximum of $4.96 \cdot 10^{-8}$. The lowest MSE was achieved using no regularization. The lowest total variation (TV, M4) resulted from applying the highest weight on the TV regularization ($\rhoTV = 8.5$), followed by the Tikhonov regularization ($\rhoT = 1.0$) and the lowest weight on the MSE ($\rhoMSE = 0.5$). The best trade-off between fidelity and variation was observed for a configuration with $\rhoMSE = 2.5$, $\rhoTV = 7.5$, and $\rhoT = 0$.

\begin{figure}
    \centering
    \begin{subfigure}{0.45\textwidth}
        \centering
        \includegraphics[width=\linewidth]{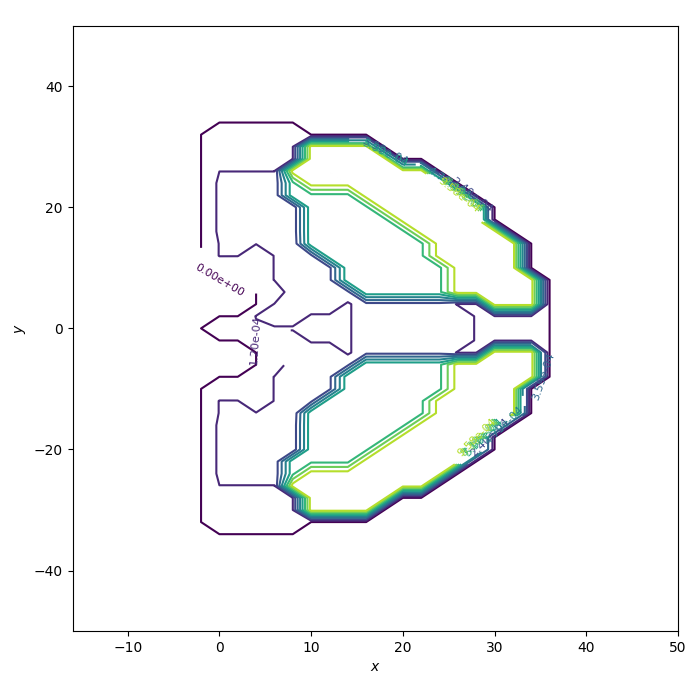}
        \caption{Occupancy distribution.\\\ }
        \label{imgApplicationAnalysisTC3OPDAndIntensityMap-original}
    \end{subfigure}
    \hfill
    \begin{subfigure}{0.45\textwidth}
        \centering
        \includegraphics[width=\linewidth]{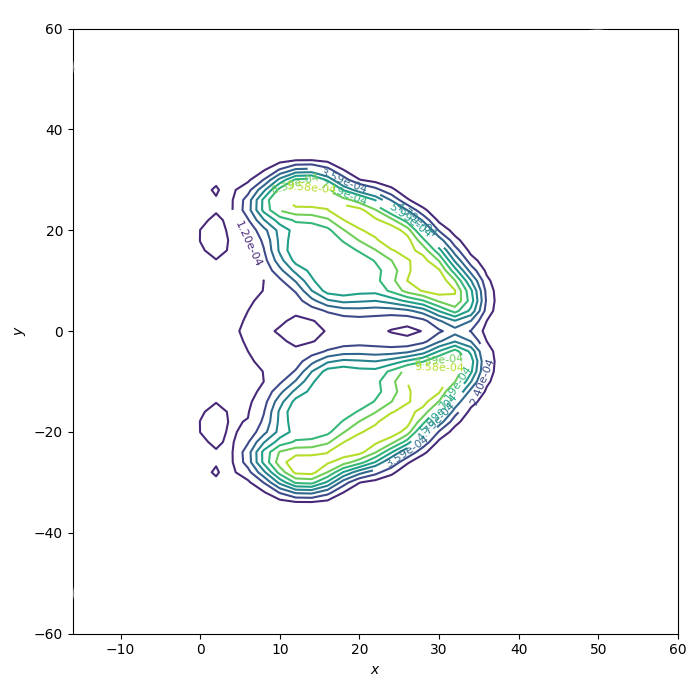}
        \caption{Intensity map with minimum MSE  ($\rhoMSE = 10$, $\rhoTV = 0$, $\rhoT = 0$).}
        \label{imgApplicationAnalysisTC3OPDAndIntensityMap-minMSE}
    \end{subfigure}
    
    \vspace{1em}
    \begin{subfigure}{0.45\textwidth}
        \centering
        \includegraphics[width=\linewidth]{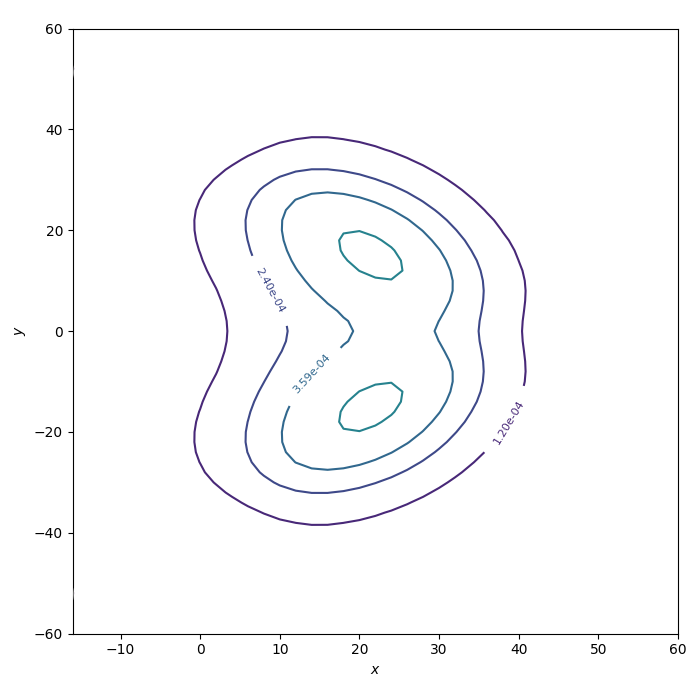}
        \caption{Intensity map with minimum TV ($\rhoMSE = 0.5$, $\rhoTV = 8.5$, $\rhoT = 1.0$).}
        \label{imgApplicationAnalysisTC3OPDAndIntensityMap-minTV}
    \end{subfigure}
    \hfill
    \begin{subfigure}{0.45\textwidth}
        \centering
        \includegraphics[width=\linewidth]{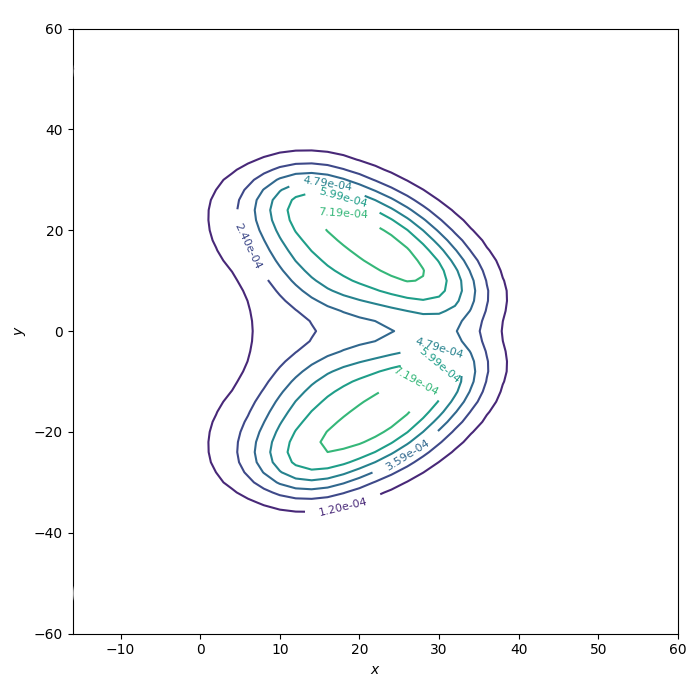}
        \caption{Intensity map with best trade-off ($\rhoMSE = 2.5$, $\rhoTV = 7.5$, $\rhoT = 0$).}
        \label{imgApplicationAnalysisTC3OPDAndIntensityMap-BestTradeOff}
    \end{subfigure}
    
    \caption{Comparison of occupancy distribution and intensity maps for TC3 under different regularization configurations.}
    \label{imgApplicationAnalysisTC3OPDAndIntensityMap}
\end{figure}

Given that the occupancy distribution is provided as a PMF, the approximation successfully smooths the rugged contour lines seen in \cref{imgApplicationAnalysisTC3OPDAndIntensityMap-original}. This is achieved while preserving the step-like transitions between plateaus -- reflected by the wider spacing between contour lines in regions where they were previously tightly packed.

The other subplots in \cref{imgApplicationAnalysisTC3OPDAndIntensityMap} show the intensity maps corresponding to the approximation 
\begin{itemize}
    \item with minimum MSE,
    \item with minimum TV, and
    \item the configuration that best balances both criteria.
\end{itemize}

The obtained Pareto front for the tested regularization weight configurations is shown in \cref{imgApplicationAnalysisOfEvaluationParetoFrontTC3}.

\subsubsection{Conclusion for Test Cases}

In cases where the MSE is already low -- such as TC1 and TC3 -- the most notable improvements are achieved by reducing variability (TV) through regularization. For TC3, this is achieved primarily via TV regularization, while TC1 and TC2 profit from a combination of TV and Tikhonov regularization.

Across all test cases, an increase in smoothness consistently leads to a decrease in surface variability (TV). This smoothness also results in lower gradients and reduced curvature, as expected from the nature of the regularization terms.

Notably, TV regularization tends to preserve sharp features or edges (i.e., regions with high gradients) more effectively than Tikhonov regularization. However, this distinction diminishes as the regularization weights increase beyond a threshold (approximately $\rhoTV, \rhoT > 2$) in the test cases studied.

Overall, the analysis does not reveal a generalizable pattern for selecting regularization weights that consistently yield an optimal trade-off between approximation fidelity and surface variability. This highlights the need for problem-specific tuning or adaptive weighting strategies in practical applications. In particular for finding the most suitable weight configuration for different instances of time.

\subsection{Optimal Control Problem} \label{secApplicationOCP}

Let $\lambda (x,y; t)$ be defined as in \cref{eqApplicationTowardsProblemOccupancyDistributionAndRiskIntensityFunction} with the obstacle set 
\[
  \Obs = \left\{\begin{bmatrix}
      10 \\ 
      -10 \\
      90^\circ
  \end{bmatrix}, \begin{bmatrix}
      20 \\
      40 \\
      -90^\circ
  \end{bmatrix}\right\}.
\]

The temporal domain is $T = \interval{t_0}{t_f}$ with $t_0 = 0$ to $t_f = 5$. All the obstacles share the same set of occupancy distributions, each regularized according to one of the previously identified weight configurations:
\begin{itemize}
    \item the one minimizing MSE,
    \item the one minimizing TV, and 
    \item the one achieving a balanced trade-off between both. 
\end{itemize}

The initial state is defined as $e_0 = \vecrow{e_{x, 0}, e_{y, 0}, 0}$, where the location vector $\vecrow{e_{x, 0}, e_{y, 0}}$ is sampled from a bivariate normal distribution $\mathcal{N}(\mu, \Sigma)$, with 
\[
  \mu = \begin{bmatrix}
      0 \\
      0
  \end{bmatrix},\quad \Sigma = \begin{bmatrix}
      1, 0 \\
      0, 1 
  \end{bmatrix}.
\]
The target state at $t_f$ is $e_f = \vecrow{60, 0, 0}$.

With the dynamics as defined in \cref{eqApplicationODESingleTrackKinematic} and the trajectory $\gamma$ of the ego vehicles state $e$ (see \cref{eqApplicationTrajectoryOfEgoVehicle}) the OCP is formulated as
\begin{subequations}
    \label{eqApplicationDescriptionEvaluationOCP}
    \begin{align}
    \underset{u}{\min} &\int_{t \in T} \iint_{\cDisk(\gamma(t))} \lambda (x, y; t) \dif x \dif y \dif t + \rho_u \norm{u}_2^2 \notag\\
  \text{s.t.}\quad &\dot{e}(t) = g(e(t), u(t)) \notag \\
    &e(t_0) = e_0 \notag\\
    &\abs{e(t_f) - e_f} \le \begin{bmatrix}
        0,
        0,
        \oo
    \end{bmatrix}^\top \label{eqApplicationDescriptionEvaluationOCP_etfInequality} \\
    &e \in W_{1, \oo}(T, \bb{R}^3) \notag \\
    &u \in L_\oo (T, \bb{R}^2). \notag
    \end{align}
\end{subequations}

The inequality constraint \cref{eqApplicationDescriptionEvaluationOCP_etfInequality} enforces convergence to the desired destination, while allowing flexibility in the final heading. The regularization weight $\rho_u$ is chosen such that the risk integral receives approximately three times the weight of the control's norm penalty. The inclusion of the control cost encourages solutions that accept low levels of risk in exchange for more effective controls.

For numerical evaluation, the OCP was discretized using 15 shooting nodes, and solved using the optimization solver \href{https://coin-or.github.io/Ipopt/}{IPOPT}. The initial guess for the state is constructed as a linear interpolation between the start and goal positions, with control inputs corresponding to a constant velocity (based on distance over $\abs{T}$) and a zero steering angle.

\subsubsection{Results for the Optimal Control Problem}
\Cref{tblApplicationAnalysisEffectOfRegularizationOnOCPM5,tblApplicationAnalysisEffectOfRegularizationOnOCPM6,tblApplicationAnalysisEffectOfRegularizationOnOCPM7} presents the statistical results obtained by solving the OCP described in \cref{eqApplicationDescriptionEvaluationOCP} for varying starting location. 

The objective function values are difficult to interpret in absolute terms, as they incorporate both the risk integral and a weighted control effort, and lack a direct empirical reference. Nonetheless, since risk dominates the objective (approx. three times the control penalty), we can interpret the values qualitatively. Approximations with regularization exhibits lower variance and a narrower min-max range, suggesting consistent behavior across different initial conditions.

Compared to the MSE-minimizing configuration with no regularization:
\begin{itemize}
    \item The number of iterations was reduced by 40\% (TV-minimizing) and 28\% (balanced), translating to average reductions of $\approx 30$ and $\approx 21$ iterations, respectively.
    \item The runtime improvement was more substantial: 50\% (TV-minimizing) and 33\% (balanced).
\end{itemize}

These results indicate that regularization -- particularly when balanced -- can significantly improve solver efficiency, likely due to the smoother, more numerically stable structure of the intensity map.

\begin{table}
    \centering
    \begin{tabular}{r|rrr}
        \toprule
         & \multicolumn{3}{c}{\makecell{\textbf{No. of Iterations}\\ (M5)}} \\
         & Min. TV & Min. MSE & Balanced \\
        \midrule
    \textbf{min} & $32.00$ & $43.00$ & $35.00$  \\
    \textbf{mean} & $45.05$ & $74.80$ & $53.75$  \\
    \textbf{std} & $21.23$ & $45.12$ & $21.90$  \\
    \textbf{median} & $39.50$ & $63.00$ & $50.00$  \\
    \textbf{max} & $124.00$ & $250.00$ & $135.00$  \\
        \bottomrule
    \end{tabular}
    \caption{No. of Iterations (M5) of the OCP solver under different regularization configurations.}
    \label{tblApplicationAnalysisEffectOfRegularizationOnOCPM5}
\end{table}

\begin{table}
    \centering
    \begin{tabular}{r|rrr}
        \toprule
         & \multicolumn{3}{c}{\makecell{\textbf{Runtime}\\ (M6)}} \\
         & Min. TV & Min. MSE & Balanced \\
        \midrule
    \textbf{min} & $ 2.38$ & $ 3.97$ & $ 3.60$  \\
    \textbf{mean} & $ 4.09$ & $ 8.23$ & $ 5.53$ \\
    \textbf{std} & $ 1.75$ & $ 4.28$ & $ 2.04$ \\
    \textbf{median} & $ 3.58$ & $ 7.16$ & $ 5.14$ \\
    \textbf{max} & $10.52$ & $25.17$ & $13.13$ \\
        \bottomrule
    \end{tabular}
    \caption{Runtime (M6) of the OCP solver under different regularization configurations.}
    \label{tblApplicationAnalysisEffectOfRegularizationOnOCPM6}
\end{table}

\begin{table}
    \centering
    \begin{tabular}{r|rrr}
        \toprule
         & \multicolumn{3}{c}{\makecell{\textbf{Objective Function Value}\\ (M7)}} \\
         & Min. TV & Min. MSE & Balanced \\
        \midrule
    \textbf{min} & $1.172\cdot 10^{-3}$ & $6.169\cdot 10^{-5}$ & $3.032\cdot 10^{-4}$ \\
    \textbf{mean} & $2.119\cdot 10^{-3}$ & $1.347\cdot 10^{-3}$ & $2.085\cdot 10^{-3}$ \\
    \textbf{std} & $6.684\cdot 10^{-4}$ & $2.055\cdot 10^{-3}$ & $1.728\cdot 10^{-3}$ \\
    \textbf{median} & $1.945\cdot 10^{-3}$ & $1.280\cdot 10^{-4}$ & $1.169\cdot 10^{-3}$ \\
    \textbf{max} & $3.271\cdot 10^{-3}$ & $6.261\cdot 10^{-3}$ & $5.086\cdot 10^{-3}$ \\
        \bottomrule
    \end{tabular}
    \caption{Objective Function Value (M7) of the OCP under different regularization configurations.}
    \label{tblApplicationAnalysisEffectOfRegularizationOnOCPM7}
\end{table}

\begin{figure}
    \centering
    \begin{subfigure}{0.45\textwidth}
        \centering
        \includegraphics[width=\linewidth]{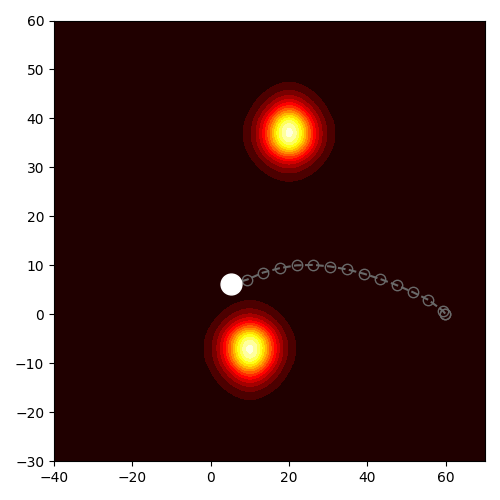}
        \caption{$t = 0.00$}
        \label{imgApplicationOCPIntensityMapTimeEvolution-step1}
    \end{subfigure}
    \hfill
    \begin{subfigure}{0.45\textwidth}
        \centering
        \includegraphics[width=\linewidth]{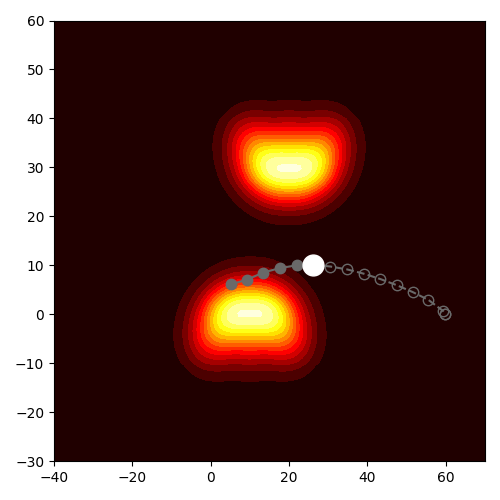}
        \caption{$t = 1.67$}
        \label{imgApplicationOCPIntensityMapTimeEvolution-step2}
    \end{subfigure}
    
    \vspace{1em}
    \begin{subfigure}{0.45\textwidth}
        \centering
        \includegraphics[width=\linewidth]{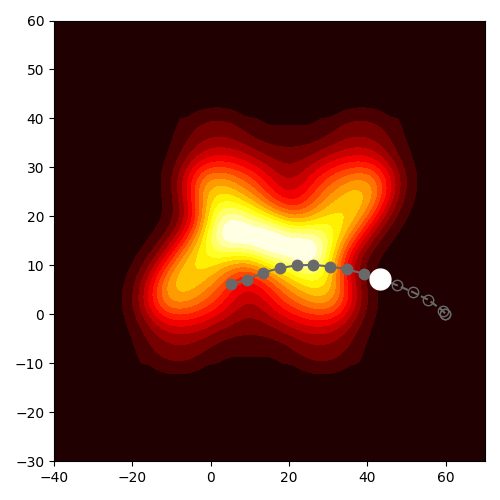}
        \caption{$t = 3.00$}
        \label{imgApplicationOCPIntensityMapTimeEvolution-step3}
    \end{subfigure}
    \hfill
    \begin{subfigure}{0.45\textwidth}
        \centering
        \includegraphics[width=\linewidth]{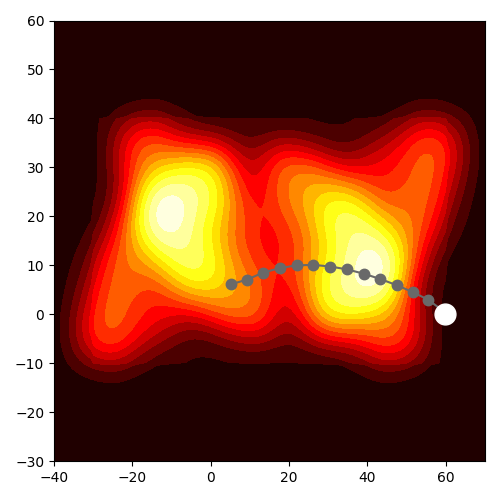}
        \caption{$t = 5.00$}
        \label{imgApplicationOCPIntensityMapTimeEvolution-step4}
    \end{subfigure}
    
    \caption{Trajectory snapshots for TC3 (balanced regularization). Intensity map $\lambda (x, y; t)$ shown as heat map.}
    \label{imgApplicationOCPIntensityMapTimeEvolution}
\end{figure}

The trajectory plots in \cref{imgApplicationOCPIntensityMapTimeEvolution} show the evolution of the OCP solution over time under the balanced regularization strategy. The smoother risk field encourages controlled steering through uncertain regions rather than strict avoidance. The white dot represents the location of ego vehicle at the specified time, while the gray, filled circle marker show historic, and the gray circle markers with no filling the future locations.

\section{Conclusion}\label{secApplicationConclusion}

In this work, we proposed a method for approximating occupancy distributions -- which describe the probability of a region being occupied by a moving obstacle -- using smooth functions that can be interpreted as intensity functions of a Poisson random field. This interpretation retains the probabilistic semantics while enabling straightforward aggregation of multiple obstacles.

The intensity maps were constructed using B-splines, placing them in a subspace of the Sobolev function space. This choice is particularly well-suited for use in objective functions of OCPs. 

Among the tested approximation weight configurations, the one that balanced the trade-off between fidelity (quantified by mean squared error, MSE) and surface variability (quantified by total variation, TV) led to a reduction in solver computational load of up to 50\% compared to configuration that only minimizes the approximation error. This performance gain is especially relevant in the context of autonomous vehicles, where computational resources are constrained. This applies particularly in model predictive control (MPC) frameworks used for high-level planning in real-time.

Besides our primary focus on applications involving OCPs, the proposed method and results are also applicable to other domains that may benefit from:
\begin{itemize}
    \item Access to derivatives (e.g., for sensitivity analysis or gradient-based planning), or
    \item The Poisson field interpretation, which simplifies the combination of multiple occupancy distributions into a coherent risk model.
\end{itemize}

\section{Outlook} \label{secApplicationOutlook}

While the proposed method has demonstrated promising results, there are several further improvements and extension. One limitation is the use of uniform knot placement, which may be suboptimal for capturing fine spatial structures. Future work could explore adaptive knot placement strategies and their effect on the OCP solver's performance. However, such adaptations introduce challenges in temporal consistency, especially when occupancy distributions are not defined for every time instance (e.g. those associated with the shooting nodes of the OCP). In that case, other methods will be required to interpolate between the intensity maps at discrete times. One viable method might be a mesh morphing technique (cf. \cite{mukundan2022}).

Another important direction involves automating the selection of regularization weights. While we have shown that particular choices improves solver performance, this balance may vary across time. An adaptive weighting scheme could yield robust, structure-aware approximations.

Finally, extending the current framework to three-dimensional spaces would make it applicable to a wider class of systems, such as aerial vehicles or drone swarms. This would require generalizing the spline-based intensity formulation and regularization to volumetric domains.

\section*{Acknowledgement}

The authors acknowledge the use of OpenAI's \href{https://chatgpt.com}{ChatGPT} (versions 4o and 4.5) to assist with grammar correction and improvements in the readability of the manuscript. No content was generated autonomously by the tool. The authors have critically reviewed and edited all AI-assisted output and accept full responsibility for the integrity, accuracy, and originality of the final content.

\section*{Funding}

This research has been conducted within dtec.bw -- Digitalization and Technology Research Center of the Bundeswehr, project MORE (Munich Mobility Research Campus). dtec.bw is funded by the European Union -- NextGenerationEU.

\bibliographystyle{tfs}
\bibliography{assets/library}

\end{document}